\documentclass[CMYK,NoSecthm]{jcomsec_author}  
\usepackage[linkcolor=black,citecolor=black,colorlinks=false,bookmarksnumbered]{hyperref} 
\usepackage{multicol}
\usepackage{enumerate}
\usepackage{amssymb}
\usepackage{multirow}
\usepackage{amsmath}
\usepackage{graphicx}
\usepackage{subcaption}
\usepackage{setspace}
\usepackage{mathtools}
\usepackage{algorithm}
\usepackage{algorithmic}
\usepackage{colortbl}
\usepackage{color}
\usepackage{ctable}
\usepackage{wrapfig}
\usepackage[numbers,sort&compress]{natbib}
\usepackage{hyperref}
\usepackage{hyphenat}
\usepackage[all]{hypcap}
\usepackage{array}
\usepackage{microtype}
\DisableLigatures[f]{encoding = *, family = * }

\newcolumntype{X}[1]{>{\centering\arraybackslash}m{#1}}

\begin{document}

\begin{frontmatter}

\title{A Sample-Based Approach to Data Quality Assessment in Spatial Databases 
with Application to Mobile Trajectory Nearest-Neighbor Search}

\author[DLS,CorAuth]{Bagher Saberi}
\ead{m.saberishahrbabaki@ec.iut.ac.ir \rm(B. Saberi)}

\author[DLS]{Nasser Ghadiri}
\ead{nghadiri@cc.iut.ac.ir \rm(N. Ghadiri)}

\address[DLS]{Department of Electrical and Computer Engineering, Isfahan University of Technology, Isfahan, 84156-83111, Iran.}

\runtitle{A Sample-Based Approach to Data Quality Assessment in Spatial \ldots}
\runauthor{ B. Saberi and N. Ghadiri} 

\corauth[CorAuth]{Corresponding author.}  
\begin{abstract}
Spatial data is playing an emerging role in new technologies such as web and mobile mapping and Geographic Information Systems (GIS). Important decisions in political, social and many other aspects of modern human life are being made using location data. Decision makers in many countries are exploiting spatial databases for collecting information, analyzing them and planning for the future. In fact, not every spatial database is suitable for this type of application. Inaccuracy, imprecision and other deficiencies are present in location data just as any other type of data and may have a negative impact on credibility of any action taken based on unrefined information. So we need a method for evaluating the quality of spatial data and separating usable data from misleading data which leads to weak decisions. On the other hand, spatial databases are usually huge in size and therefore working with this type of data has a negative impact on efficiency. To improve the efficiency of working with spatial big data, we need a method for shrinking the volume of data. Sampling is one of these methods, but its negative effects on the quality of data are inevitable. In this paper we are trying to show and assess this change in quality of spatial data that is a consequence of sampling. We used this approach for evaluating the quality of sampled spatial data related to mobile user trajectories in China which are available in a well-known spatial database. The results show that sample-based control of data quality will increase the query performance significantly, without losing too much accuracy. Based on these results some future improvements are pointed out which will help to process location-based queries faster than before and to make more accurate location-based decisions in limited times.
\end{abstract}

\end{frontmatter}
\addtolength{\parskip}{2mm}


\section{Introduction}
In the last decade, we witnessed the public use of spatial data. Google Earth is a good example which is becoming increasingly popular among the general public as well as being used by specialists to develop new applications. Spatial data and spatial reasoning are playing an increasingly important role in today's advanced computing demands such as context-aware systems, pervasive computing and decision support systems \cite{GH}. Geographic information is a prerequisite for many specific fields and collecting geographic data needs a lot of labor, resources and money. Moreover, it is impossible to re-collect a part of the geographic data. Therefore, guaranteeing the quality of spatial data is of great importance in enhancing the scientific aspect of the relevant decision-making \cite{devillers2006fundamentals}. On the other hand, huge increase in the volume of spatial data, the routine but complex nature of spatial operations such as distance computation and topological reasoning as well as provision of non-expert driven computer information systems have resulted in vastly increased potential for error and misleading information in spatial data products \cite{xie2008quality}. 
Assessing and reporting quality of spatial big data are very important aspects of producing and using this type of information. Quality assessing methods have a great impact on the efficiency and usability of data. Although there are many methods for such tasks but we tried to find a way that has a better efficiency for the emerging field of spatial big data.

The sample-based method, regarding its static or periodic nature, can lead to a great efficiency improvement in spatial data quality assessment algorithms. In this approach quality of data sources is estimated by sampling the base data sources and using the quality of these samples to assess quality of any information product.

The primary contributions of this paper are: (1) introducing the sample-based spatial data quality assessment method, (2) evaluating the effect of sampling on the quality of spatial data, and (3) improving the performance of trajectory nearest-neighbor search controlled by the proposed data quality assessment method. The paper yields insights as to how quality estimates for the base data sources can be used to provide quality estimates for information products generated from them. Thus, it is neither necessary nor useful to consume a lot of resources to inspect entire databases.

The rest of the paper is organized as follows: Section~\ref{sec:bascon} explains the basic concepts of data quality and spatial data quality. Section~\ref{sec:proposedmethod} develops our proposed method. Section~\ref{sec:experimentresults} introduces data sources, tools that we used in this study, experiments and results. Section~\ref{sec:openissues} points out some open research areas in this context and Section~\ref{sec:conclusion} concludes the paper.
\section{Basic Concepts}
\label{sec:bascon}
In recent years data quality has gained more and more importance in science and practice. This is mainly due to an extended use of data warehouse systems, cooperative information systems and a higher relevance of customer relationship management. In this section, we give an overview of data quality assessment in general, followed by more specific field of spatial data quality assessment. The concept of trajectory nearest-neighbor query is also introduced in last part as the application domain that we used for evaluation of our proposed spatial data quality assessment approach.
\subsection{Data Quality}
In order to assess data quality, first of all a clear picture of data quality is needed. Many researchers noticed that the benefits of data depend heavily on completeness, correctness, consistency and timeliness. These properties are known as data quality (DQ) dimensions. One of the major causes for the failure of information systems to deliver value can be attributed to data quality. The major challenge of data quality research is to define data quality from the consumer's point of view in terms of fitness for use and to identify dimensions of data quality according to that definition. More than a hundred of data quality dimensions were uncovered in early research works, but many of these dimensions are not much useful in many application areas. Some of the very important measures of data quality are relevance, accuracy, timeliness, accessibility, utility, interpretability, completeness, coherence and comparability.

On the other hand, we need a method to assess data quality and then deliver the results to the user. Quality reporting is about providing the reports that contain information about the quality of data. A quality report provides information on the major quality attributes of an information product so that the user would be able to evaluate product quality. In the optimal case, quality reports are based on quality indicators. Metadata is considered to be the main quality report method but many other methods have been proposed \cite{goodchild2007beyond,Borek2011Classification}\cite{boin2007communicates,madnick2009overview}\cite{woodall2010hybrid,peukert2011amc}.
A data quality assessment template is usually required that specifies the information each reviewer should assemble for the target data system. The template organizes the metadata and other information, by data system and assessment process. It also serves as an outline for the assessment report.

There are many quality assessment methods. Cross-domain analysis can be applied to data integration scenarios with dozens of source systems. It enables the identification of redundant data across tables from different, and in some cases even the same, sources. Data validation algorithms verify if a value or a set of values is found in a reference data set. Domain analysis can be applied to check if a specific data value is within a certain domain of values. Matching algorithms are used to identify duplicates such as two customer records that refer to the same customer and many other methods that are currently in use \cite{Borek2011Classification,dustdar2012quality}.
\subsection{Spatial Data Quality}
The context within which geospatial data are used has changed significantly during the past ten years. Users have now easier access to geospatial data but typically have less knowledge of the geographical information domain. So they have limited knowledge of the risk related to the use of low quality geospatial data. Moreover, spatial data and many spatial data products are being used today for making very important decisions that may have very strong impact on people’s lives. Using spatial data with low quality may even have legal complexities and therefore liability is another aspect of data quality \cite{onsrud2009liability}.

Many methods for assessing the quality of spatial data have been proposed over the years \cite{talhofer2011spatial,Talhofer2009Geospatial}\cite{laxmaiah2013conceptual,droj2011}\cite{jakobsson2011,marquez2012methodological}. In general, these methods are classified in two major groups. (1) Methods that are based on internal quality. Internal quality corresponds to the level of similarity that exists between the data produced and the ``perfect'' data that should have been produced. In practice, the evaluation of internal quality does not use the perfect data that has no real physical existence since it is an ``ideal'' dataset, but uses a dataset of greater accuracy than the data produced, which is called ``control data'' or ``reference data''. (2) Methods that use external quality which is defined as the level of concordance between data and the user requirements. External quality implies that quality is a relative concept and the same data can be of different quality to different users. While some methods exist to evaluate internal data quality, the evaluation of external quality remains a field that has not been much explored.
A large number of data quality assessment approaches use artificial intelligence and expert systems \cite{devillers2007towards,ballou2006sample}\cite{duckham2000spatial,mostafavi2004ontology}\cite{zargar2009operation,vizhi2012data}\cite{Henriksson2007Ontology,hunter1999new}. However, efficiency is not the primary concern in most of such sophisticated methods. Meta data is almost the most common method for data quality assessment and dissemination.

Among the quality assessment methods, the sample-based approach can offer a huge impact on efficiency. This is a new approach that almost has been neglected in the past for spatial data quality, but with a good design it can be very useful. The large amount of spatial data leads to complexity of data quality measurement, which in turn will cause performance impairment. To solve this problem, data quality control based on sampling can be used. In other words, we can use a carefully selected subset of our big dataset to assess its quality. However, shrinking the volume of spatial data and at the same time preserving the quality, can be very important. For example accuracy, completeness and topological consistency can be lost when we apply a sampling method on a spatial dataset. So this method should be used very carefully and samples should be selected precisely.
\subsection{Trajectory Nearest-Neighbor Search}
To evaluate the proposed spatial data quality assessment approach, we will apply it to an emerging application domain, namely mobile users’ trajectory nearest-neighbor search. Finding the nearest-neighbor is a common operation in spatial data context. Clearly we have a spatial object, like a point or a route, and we want to find an object in a collection with the minimum distance. Some spatial objects, like routes, can be very large in terms of data records. It means that they could consist of so many spatial points that spatial operations like finding the nearest-neighbor and its distance could be very time consuming. So if we can find a way to shrink the size of the routes and, at the same time, preserve the quality of data, we would improve the efficiency without impairing our reasoning and decision making \cite{DBLP:conf/gis/ChenJZLY09}.
The essence of this study is conducting a well-known spatial operation, i.e. trajectory nearest-neighbor search, using the complete spatial datasets and objects and then sampling the same data sources, performing the same operation and evaluating its effect on the quality of results and on the performance. Sampling is carried out at different rates so we can observe the effect of this parameter. Using this method, decision makers can determine whether the quality of spatial data after sampling meets their needs or not.
We applied our method on a dataset of mobile user GPS trajectories as a spatial data source. Then we performed a particular spatial operation on them ( finding the nearest-neighbors ). Then the same routine was applied using the sampled data. The experiments in this paper will show the change in quality level of these spatial data after sampling. 
\section{The Proposed Method: Sample-Based Spatial Data Quality}
\label{sec:proposedmethod}
The main idea in our approach is to take samples from the original data, then evaluate the quality of the sampled data, and use the attained information to estimate the quality of any information product that can be derived from the original data. Sampling is carried out only once or on some predetermined periodic basis. The major advantage of this approach is that only the base data need to be sampled. The quality of data or the number of identified deficiencies can be context-dependent. Thus, the quality measure used for a given dataset will vary according to its use.

The information products that can be derived from original data are the result of some combination of algebraic operations. Therefore, the quality of any information product can be calculated by applying some algebraic operations on the quality of base data. In fact, the sampling method can also be used for measuring the quality of data, even if we have multiple tables of data as sources. Every data that might be requested from a database is a result of a set of operations like restriction, Cartesian product, union and projection on basic tables of data. So we can calculate the quality of every information product if we have a view of quality of the base tables \cite{devillers2007towards}. Furthermore, this idea is extendable to multiple independent databases and data integration \cite{ballou2006sample}.  This could also be a wide area for future research on spatial big data. 
For example if we should apply some constraints on a table in order to get the desired data  units, then we use the samples taken from the source table to assess the quality of information product \cite{devillers2007towards}. Or in the case of union of two tables, suppose that no duplicates exists and the first table has N data units and P1 is the estimate for the acceptable data units and n1 is the size of sample and M, P2 and n2 are corresponding factors for the second table. An estimate of the fraction of acceptable data units in the union of two tables can be calculated using Eq. (1) below \cite{devillers2007towards}:
$$P = (n1 * P1 + n2 *P2) / (n1 + n2)                             (1)$$ 
Determining the quality of an information product when there are duplicate data units is considerably more difficult but still possible. 
it's possible to say that for all of the other algebraic operations, like join, set difference and product, we can obtain an estimate of the quality of information product by applying some operations on the assessed quality of source data. This is a very useful property of the sample-based approach in the context of spatial data.

Under this circumstance, a large enough sample must be taken so that defective records also appear in the sample \cite{devillers2007towards}. A standard rule of thumb is that the sample should be large enough so that the expected number of defective items is at least two. Since sampling with replacement is a binomial process, then n, the size of the sample, must satisfy the inequality of Eq. (2). \cite{ballou2006sample}:
$$n ≥ 2 / (1 – P)                                                     (2)$$
Where P represents the true proportion of acceptable data units. Clearly, there needs to be some estimate for the value of P in order to use this inequality. One way of estimating P is by taking a preliminary sample before initiating the first round of sampling \cite{levy2013sampling}. So we will apply this sampling approach to real data sets followed by evaluating the effect of the proposed method in a real application.
\section{Experimental Results}
\label{sec:experimentresults}
In this section, we present the results of applying our proposed sample-based spatial data quality assessment as a control mechanism for efficient processing of trajectory nearest-neighbor search
\subsection{Datasets}
For this case study we used 36 mobile phone trajectories from Microsoft GeoLife  as our source of base data. These trajectories are obtained from different users, in different times, and in different sizes so we can neglect the effect of these factors. Each one of these trajectories shows a path that a user has followed during an arbitrary time of day and hence includes a number of spatial points and each point is consisted of its spatial attributes like latitude, longitude and altitude. It also contains some collateral information like time and date. The size of trajectories varies from very small, containing about 25 points, to very large trajectories that include tens of thousands of points.

\begin{figure}[ht]
\centering
\includegraphics[scale=0.25]{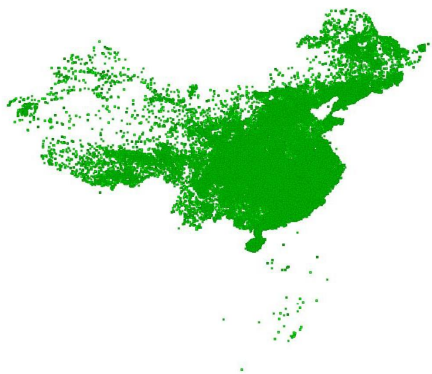}
\caption{China in GeoNames database}
\label{fig:china}
\end{figure}

\begin{figure}[ht]
\centering
\includegraphics[scale=0.25]{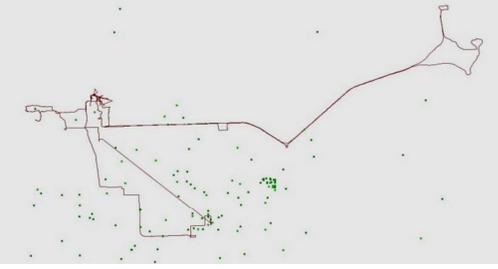}
\caption{A GeoLife trajectory and its neighbors in GeoNames database}
\label{fig:GeoTraj}
\end{figure}

The GeoNames  database is another geospatial database that provides useful spatial data for almost all countries including longitude, altitude and latitude for points of interest, population and administrative division of important points, like cities, rivers, lakes and many other geographical entities. For this paper we took a part of source data from this database. Figure~\ref{fig:china} and Figure~\ref{fig:GeoTraj} show a sample of these two databases.  
\subsection{Experiments}
\begin{figure}[ht]
\centering
\includegraphics[scale=0.17]{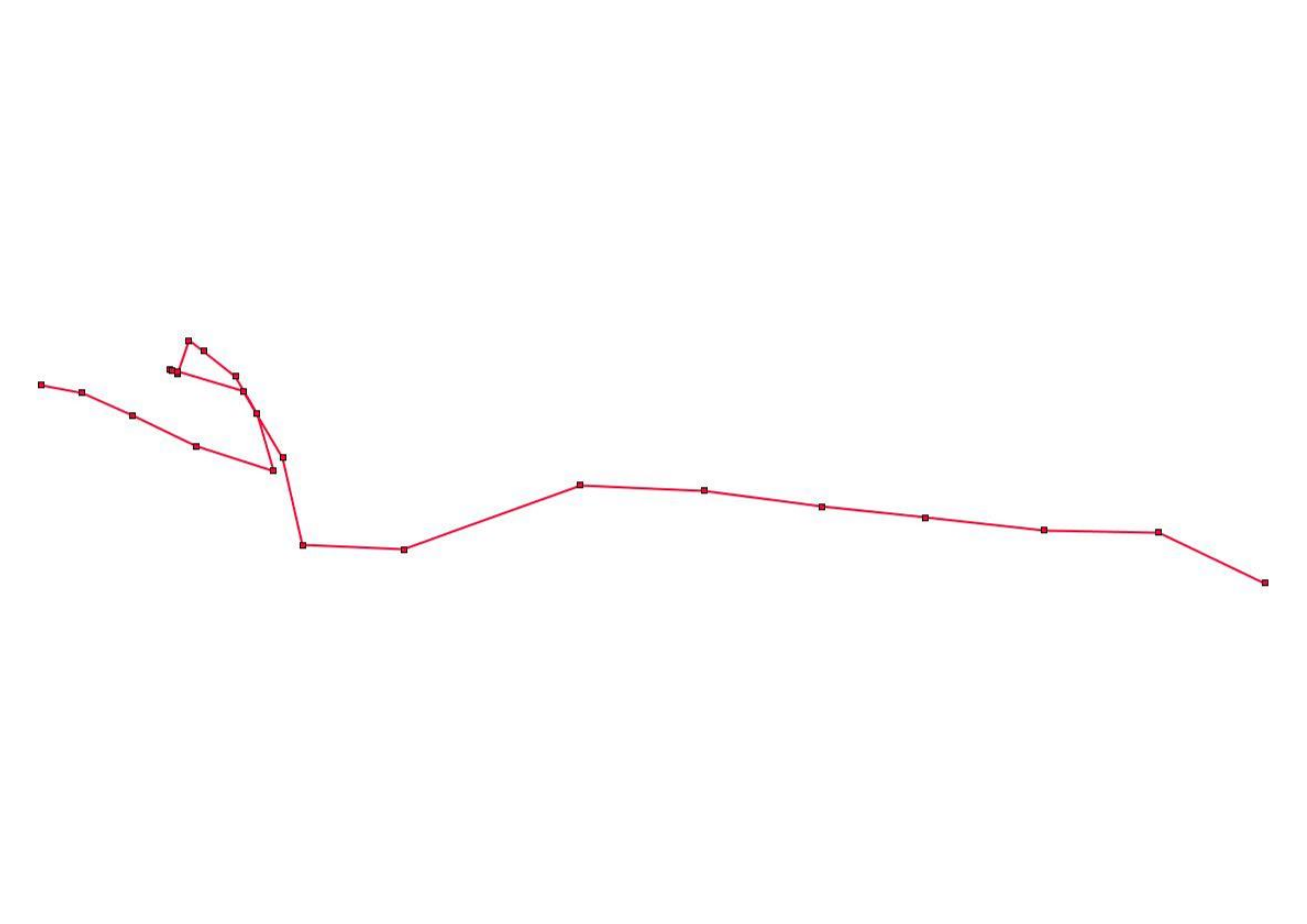}
\caption{Complete trajectory}
\label{fig:comtraj}
\end{figure}

\begin{figure}[ht]
\centering
\includegraphics[scale=0.25]{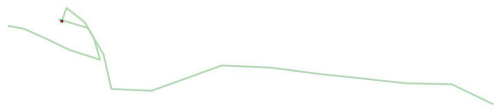}
\caption{Sampled route with rates 1/15, 1/20, 1/30.}
\label{fig:samtraj}
\end{figure}

As mentioned in Section 3, a sampling algorithm is of great importance for spatial data quality assessment and serves as a promising research area \cite{hunter2005next,devillers2010thirty}\cite{DBLP:journals/tgis/Guptill08}. In this paper we used the aforementioned periodic method for sampling, because efficiency is our primary concern and the static nature of periodic sampling can improve it . In the first step of our experiments, our sampling rate is 1/2. The sampling procedure was applied with another six different sampling rates including 1/3, 1/5, 1/10, 1/15, 1/20 and 1/30, to all 36 trajectories. Figures~\ref{fig:comtraj}  to ~\ref{fig:samtraj} depict the smallest trajectory in its complete and sampled forms visually.

For our study we used a set of spatial data quality metrics namely positional and attribute accuracy, precision and logical consistency to examine the quality. In the sample-based approach all of spatial data quality measures are affected. For instance if a spatial entity is up to date, its samples may be not. On the other hand, if an entity is topologically consistent, its samples may be inconsistent, because it does not include all of its original data. As for the updateness factor, since some of the information is lost in the sampling process, updateness is affected too.
Sampling rate in this context is also an open area for research. For example, if the actual error rate is 1 percent and a sample size of 10 is taken, then only occasionally an error will show up in the sample.

We tried to estimate the effect of sampling on the quality of spatial data and operations and demonstrated the results. First we found the nearest-neighbor to the complete trajectory and measured the spheroid distance between our trajectory and its nearest-neighbor. Then we took samples from trajectories with different sampling rates and followed the same procedure to observe the effect of different sampling rates on the quality of finding nearest-neighbor and its distance.

\begin{figure}[ht]
\centering
\includegraphics[scale=0.25]{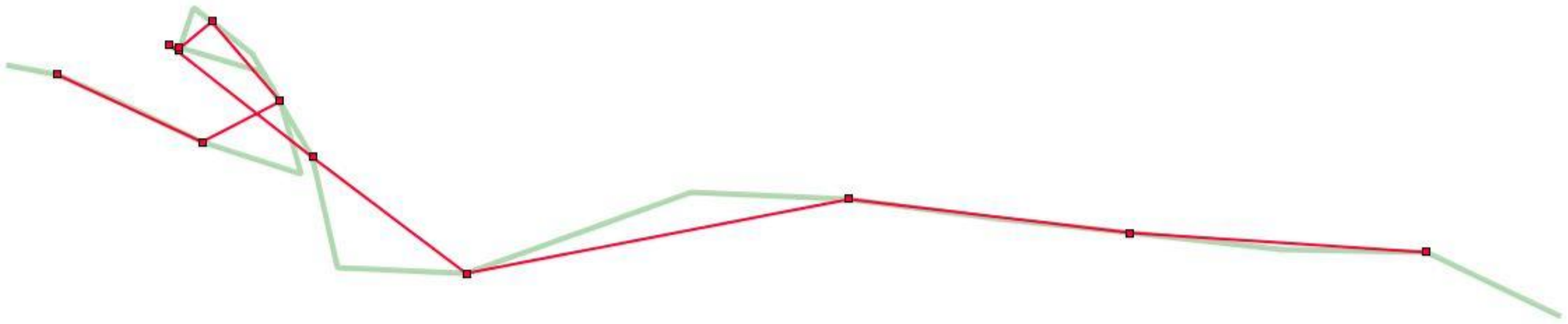}
\caption{Sampled route with rate 1/2.}
\end{figure}

\begin{figure}[ht]
\centering
\includegraphics[scale=0.25]{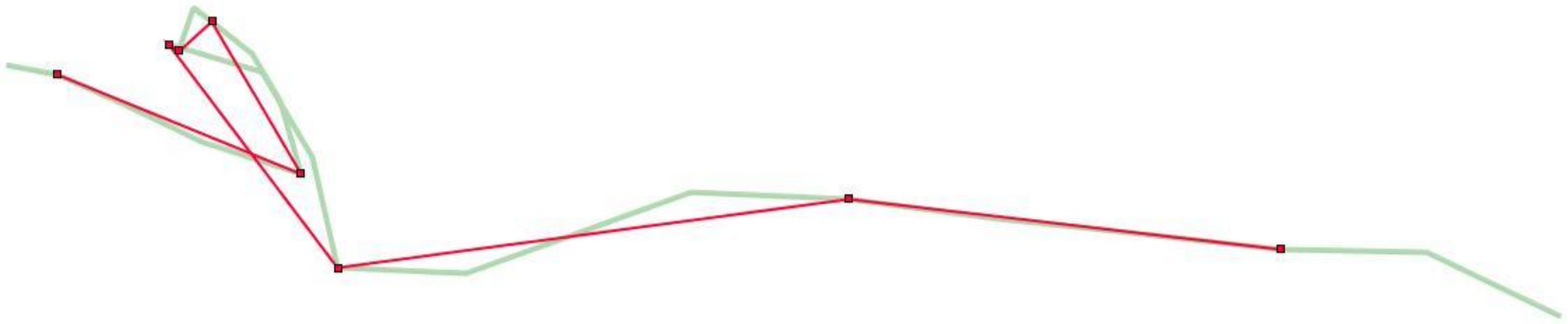}
\caption{Sampled route with rate 1/3.}
\end{figure}

\begin{figure}[ht]
\centering
\includegraphics[scale=0.25]{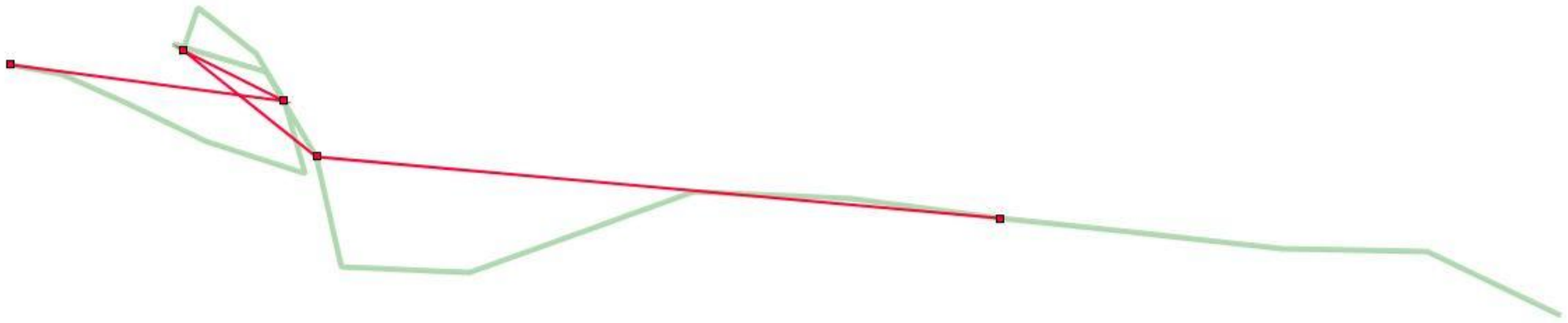}
\caption{Sampled route with rate 1/4.}
\end{figure}

\begin{figure}[ht]
\centering
\includegraphics[scale=0.25]{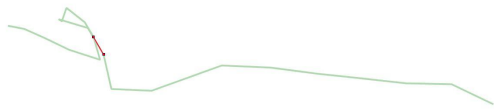}
\caption{Sampled route with rate 1/10.}
\end{figure}

At first we applied the sample-based approach to datasets to find out whether they are of good quality and suitable for use or not. This is essentially the meaning of fitness for use quality measure. According to this measure a dataset that is of good quality for a particular user may be not so good for another one. Actually this measure is so important that over the years many methods have been proposed to somehow quantize it \cite{hunter1999risk}. Therefore we decided to check whether this database is good for this purpose or not.

In order to specify whether data available in GeoNames is of good quality, we made a dataset including about 120 records and then a 20-record uniformly distributed random sample was taken. We chose this size so that at least two errors appear in the sample. Then we measured positional accuracy according to the internal quality measure, or by comparing longitude and latitude in our samples to a reference dataset captured from many other credible sources including Wikipedia and Google Earth. For the cities if the error was under 0.01 we considered the record as acceptable. For the other points the error must be under 0.001 in order to be viable.

\begin{figure}[ht]
\centering
\includegraphics[scale=0.25]{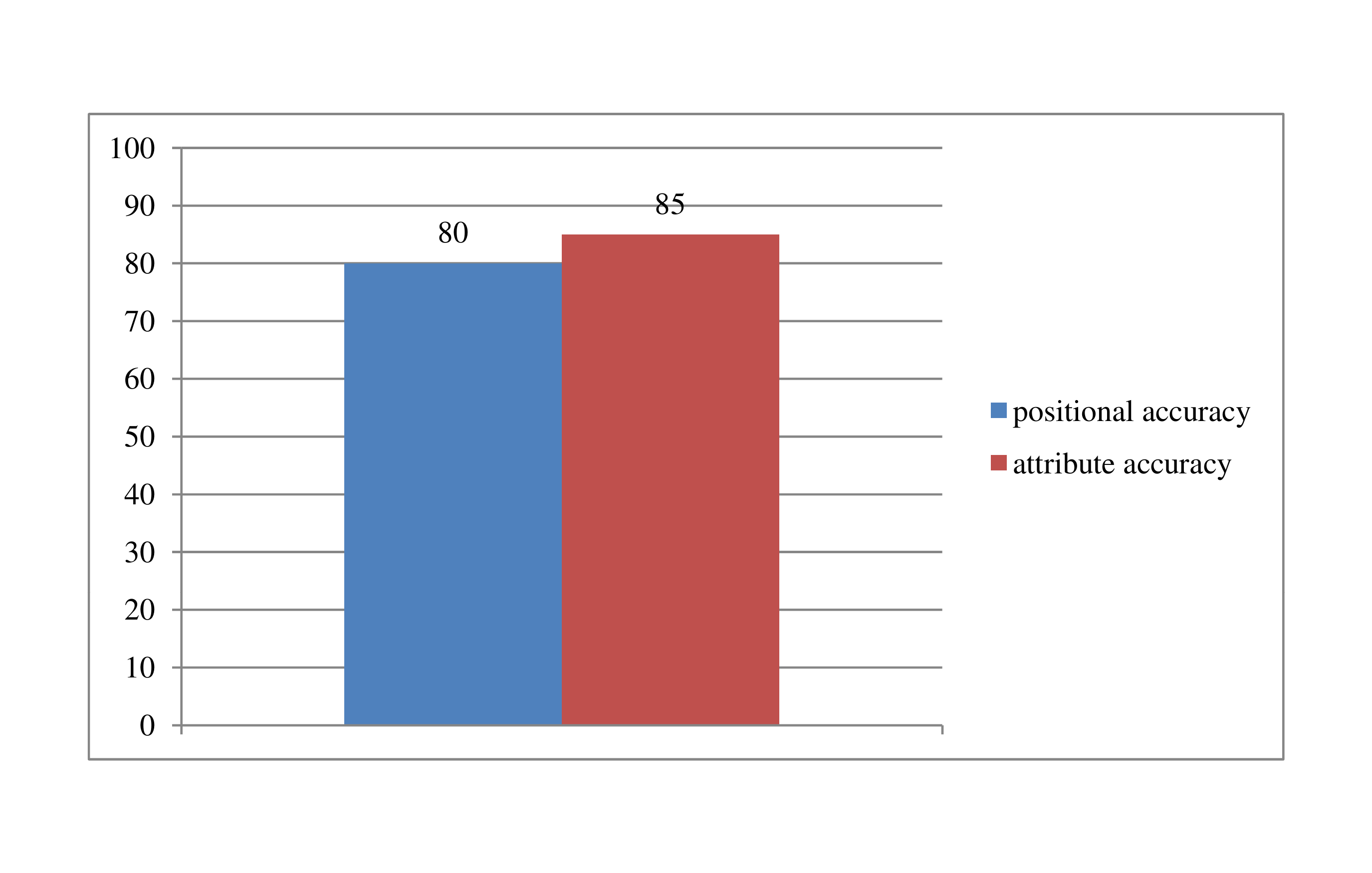}
\caption{Results of the first round of quality assessment}
\label{fig:firstasses}
\end{figure}

The threshold for accepting or rejecting a sample is very important and could be the subject of further research. In this work we considered 90 percent and 10 percent to accept or reject a sample. Figure~\ref{fig:firstasses} demonstrates the results for the first sample.

Based on our standard method, we needed a second sample. In the second round of our experiment, we took a 30-record random sample and measured two aforementioned quality factors. The second sample is also a uniformly distributed random sample.

\begin{figure}[ht]
\centering
\includegraphics[scale=0.25]{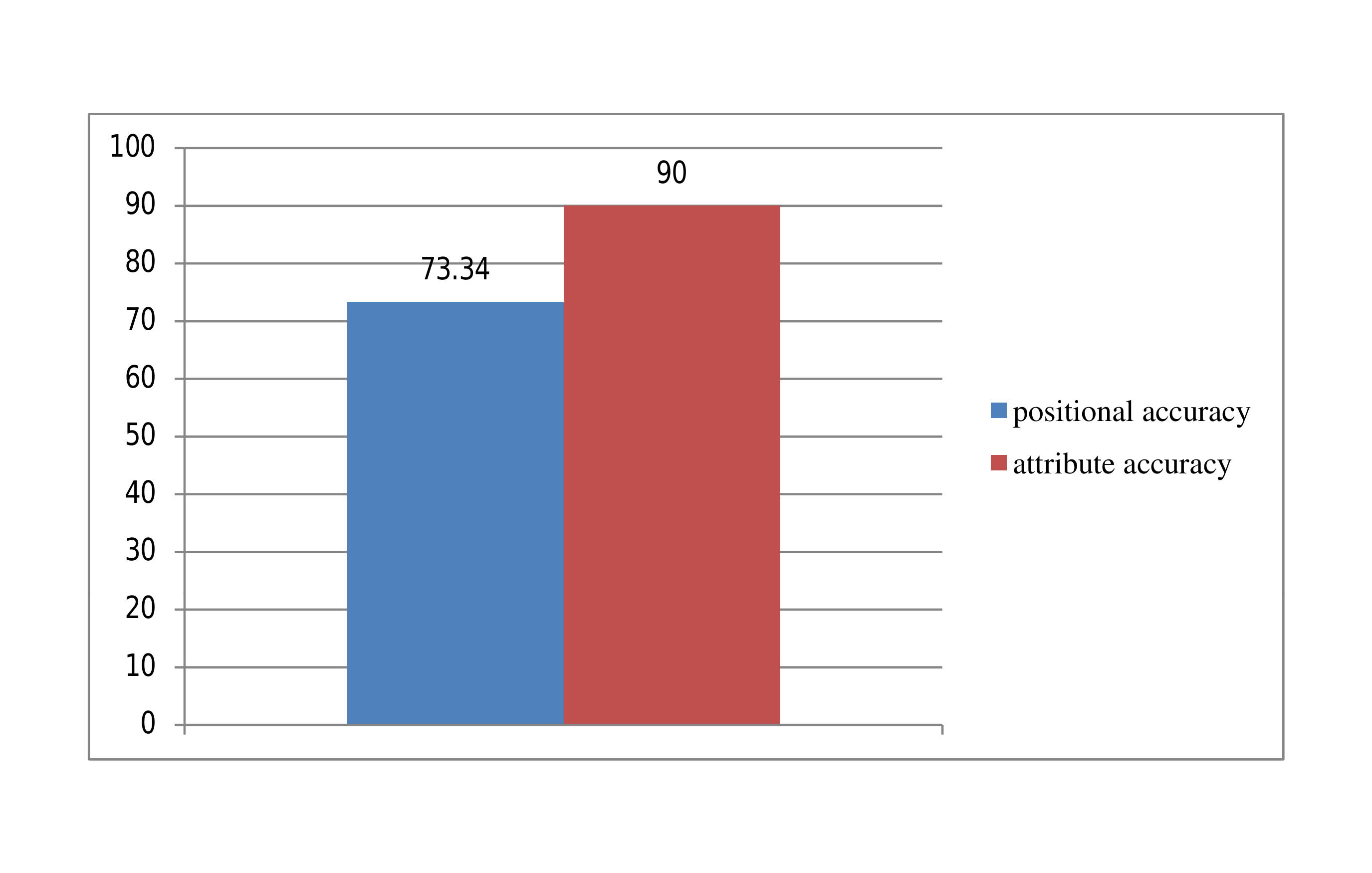}
\caption{Results of the second round of quality assessment}
\label{fig:secondasses}
\end{figure}

The results of the second round of quality assessment for the second sample are depicted in Figure~\ref{fig:secondasses}. The method of measuring the accuracy of positional data is same for the first and second round and is based on comparison with the reference dataset acquired from credible databases.

One of the important quality metrics of spatial data is updateness. In GeoNames database the date of last data update is available. Users know how old data are and they can choose to use them or not.
it's predictable that there is not any pattern in updateness of data. This is understandable because GeoNames provides possibility of entering or editing data for the users.

Suppose that a user requires the longitude and latitude of the cities and their hotels in both Iran and France. In GeoNames this data is stored in two separate tables. This information product can be derived from applying the union operation on two tables. Hence using sample-based approach we can present an estimate of data quality for an information product in an efficient way.

Recall from the previous section that we measured the positional accuracy of data for Iran to be about 73.34 percent. In a similar way, the positional accuracy of data for France is 86.6 percent. So according to Eq. (2), the positional accuracy of information products can be calculated as:$$(30 * 73.34 + 30 * 86.6) / (30 + 30) = 0.799$$
Therefore, a user can decide if the quality of data of every required information product is enough or not.
After assessing the quality of our data sources now we can examine quality of operations on sampled data and compare it with operation on complete data. In other words, we want to know that if we performed the same operation on sampled data instead of complete data, like a sampled route instead of complete route, how much error would our spatial results have. For instance, if we measure the distance between a particular point and a complete route, and then measure the same quantity but with the sampled route in different sampling rates, how different the results will be. Spatial operation in this study is finding distance. First we calculated the distance between a complete trajectory and its nearest-neighbor. The database environment for this job was PostGIS 1.5 and our spatial data visualization was done by OpenJUMP software. The necessary  code for constructing the database and necessary conversions are in SQL format. The same procedure was then applied to the trajectories and we loaded 36 trajectories to a database. The code for this task is also in SQL. In the next step, since each trajectory was a sequence of spatial points, we had to convert them to a spatial route. At this point, our data is ready for performing our experiments. The process of finding the trajectory's nearest-neighbor and their distance forms the main part of this experiment. This task was done using PostGIS spatial operations. 

\begin{figure}[ht]
\centering
\includegraphics[scale=0.45]{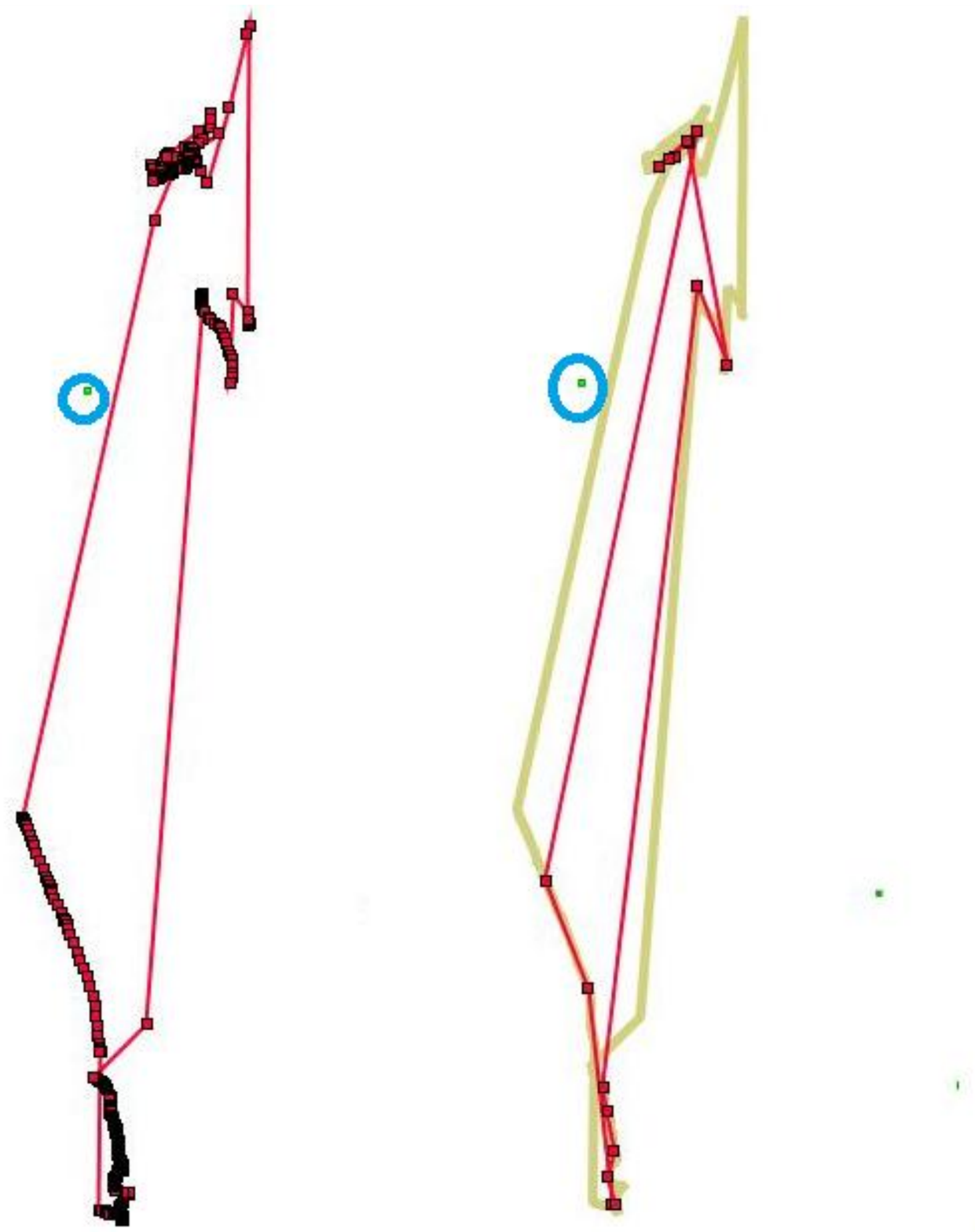}
\caption{complete and sampled trajectory and its nearest-neighbor}
\label{fig:comne}
\end{figure}

The final part of our study includes sampling trajectories with different sampling rates, as well as converting these samples to routes and ultimately finding the distance between the sampled routes and their nearest-neighbor. All of the programs that we used in this study are also in SQL format. It's clear that by sampling the trajectories the quality and particularly the accuracy of calculated distance between these trajectories and their nearest-neighbor may be affected. Figure~\ref{fig:comne} shows this important aspect by showing a complete trajectory and its nearest-neighbor as well as the same trajectory in sampled form and it's computed nearest-neighbor . The process of finding nearest-neighbor was performed using PostGIS spatial function STDistance(), which returns the spheroid  minimum distance between two geographic points in meters. For maintaining integrity, all of the lengths in this study were measured using spheroid system so that our results would be more realistic. Also all of our measurements are in the metric system and we converted all of these results to kilometers so as to compare them together. Though, all of the numbers were in a very good precision and we measured the lengths with 12 decimal points.

\begin{figure}[ht]
\centering
\includegraphics[scale=0.25]{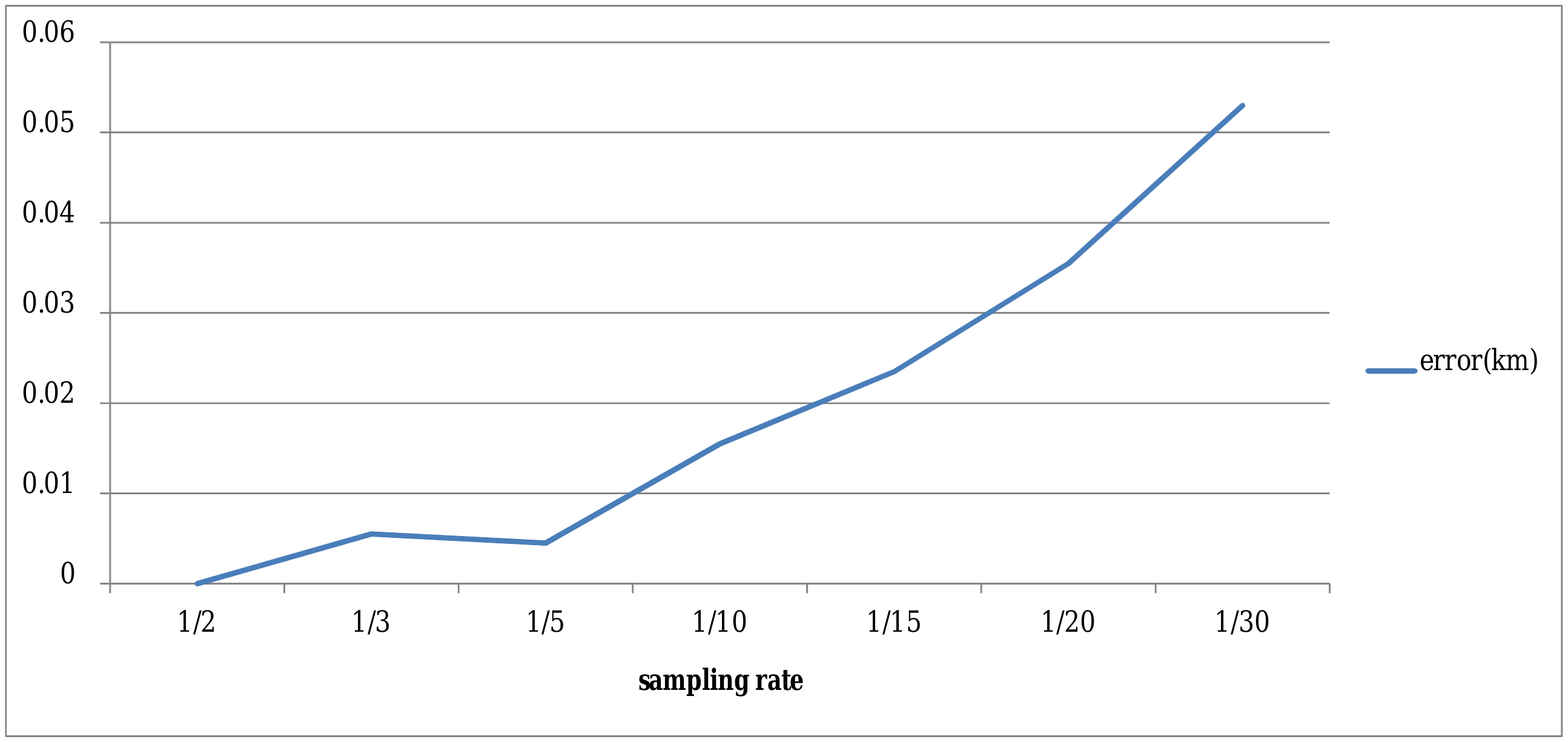}
\caption{Mean error for all trajectories}
\label{fig:meanerorall}
\end{figure}

By conducting the previously explained steps, we observed the following results. First it's obvious that with an increase in the sampling rate, error in the calculated distance increases too. Figure~\ref{fig:meanerorall} shows that the mean error confirms this.

\begin{figure}[ht]
\centering
\includegraphics[scale=0.25]{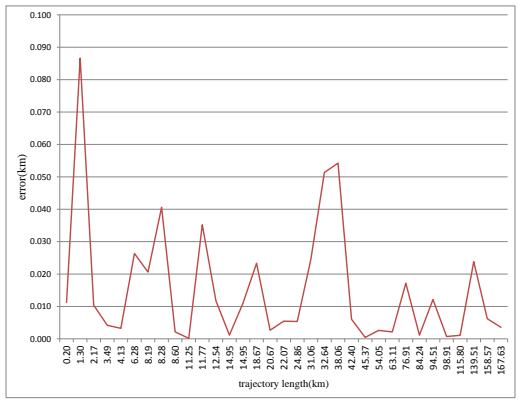}
\caption{Mean error}
\label{fig:meanerror}
\end{figure}

Figure~\ref{fig:meanerror} shows the mean error for all trajectories. You can see that mean error is changing in a narrower band for larger trajectories. This means that you may apply sampling on larger trajectories and get more accurate results with possibly huge performance gain. If the trajectories are too short, with a little change in the length of route, you may get a huge change in the error size. 

\begin{figure}[ht]
\centering
\includegraphics[scale=0.25]{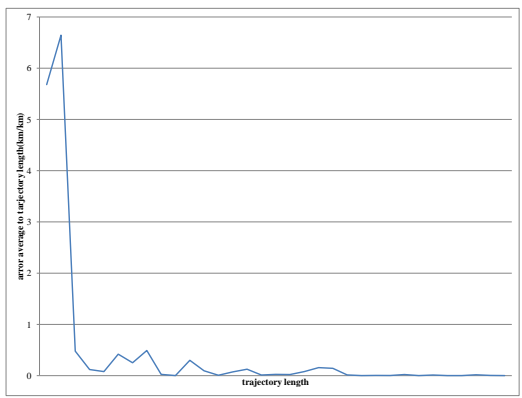}
\caption{Mean error to trajectory length}
\label{fig:meanerrorlength}
\end{figure}

One can observe from Figure~\ref{fig:meanerrorlength} that for longer trajectories, the ratio of error size to the trajectory length is lower, so for very long trajectories you can neglect the sampling error. 

\begin{figure}[ht]
\centering
\includegraphics[scale=0.25]{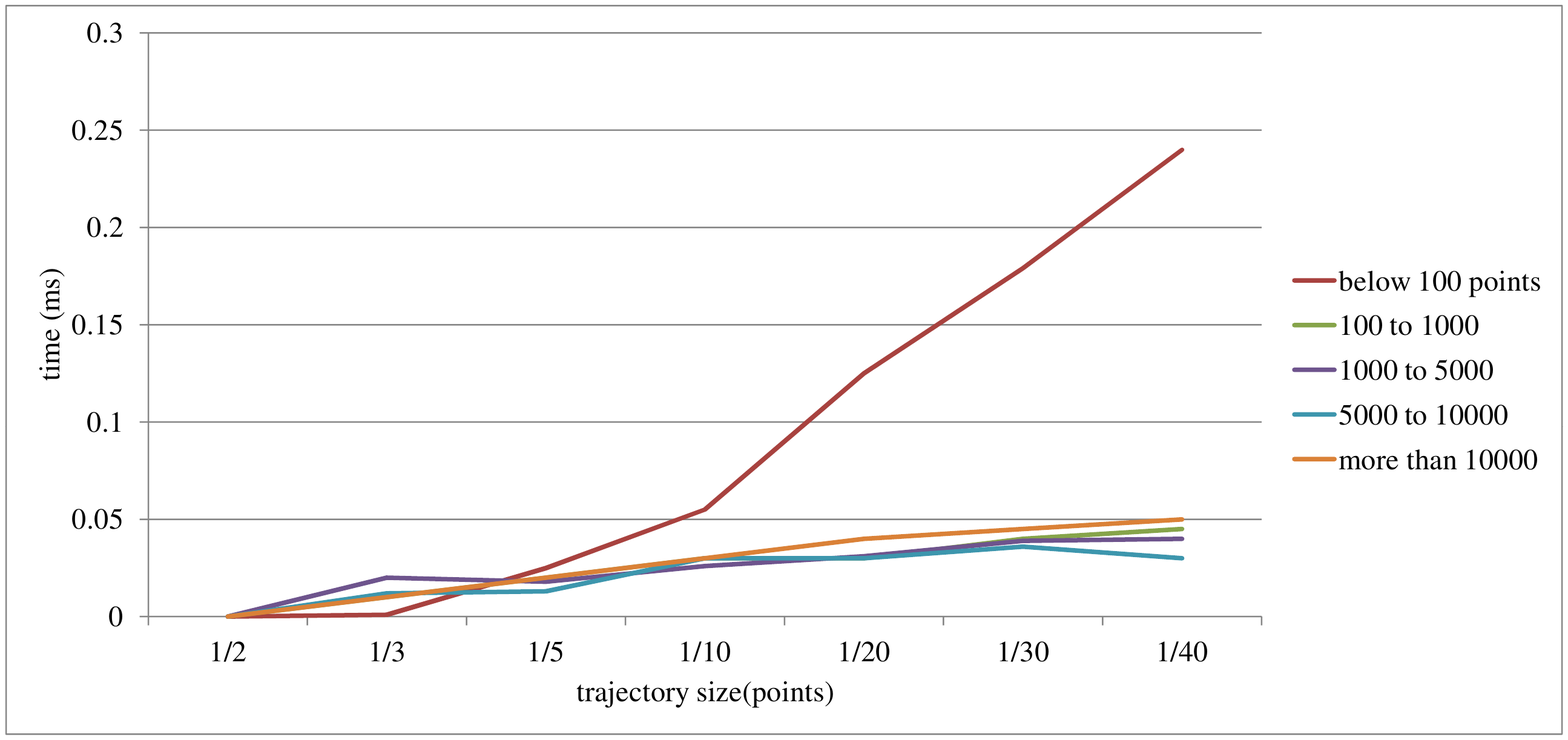}
\caption{Error size for trajectories}
\label{fig:errorsize}
\end{figure}

As for the error caused by sampling rate, it's more complicated. It can be seen from Figure~\ref{fig:errorsize} that for smaller trajectories the absolute size of error increases very faster and therefore for the small trajectories with low number of points, it is not very reasonable to apply sampling. However, based on our results, this method can be used for longer trajectories.

In terms of other quality factors, the completeness is decreased by sampling as expected. Assuming that we have a complete dataset as our source, by sampling we are actually deleting useful information. So our resulting database may not be much complete.

\begin{figure}[ht]
\centering
\includegraphics[scale=0.25]{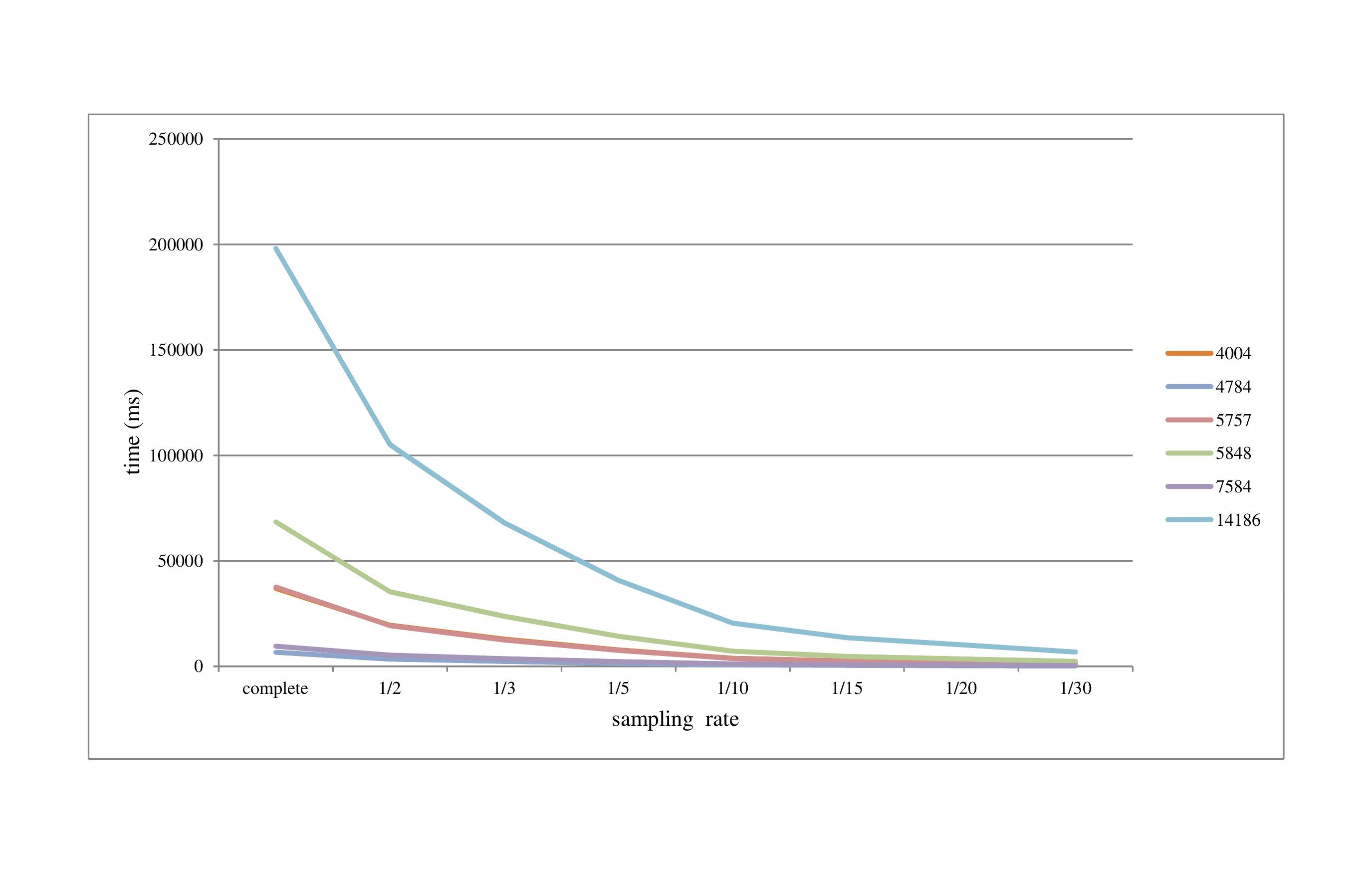}
\caption{Execution time of finding nearest-neighbor for large trajectories with more than 3000 points}
\label{fig:exetimelarge}
\end{figure}

\begin{figure}[ht]
\centering
\includegraphics[scale=0.25]{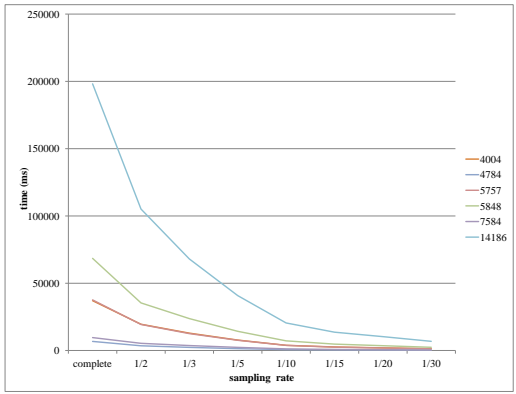}
\caption{Execution time of finding nearest-neighbor for small trajectories with less than 1000 points}
\label{fig:exetimesmall}
\end{figure}

From the performance point of view, the results are interesting. Figure~\ref{fig:exetimelarge} shows the execution time of finding nearest-neighbor for larger trajectories and in Figure~\ref{fig:exetimesmall} for smaller ones. In both configurations each line indicates the execution time for a trajectory and the legend shows the number of spatial points in it. It can be observed that for all trajectories, being large or small, the execution time of finding nearest-neighbor is proportional to the size of the sample. If we consider the trade-off between the execution time gain and the loss of quality it seems that with large and medium size the sampling method could yield better results. This is due to the fact that when we decrease the size of the sample for larger trajectories to obtain a huge gain in execution time, the quality loss is reasonable .

\begin{figure}[ht]
\centering
\includegraphics[scale=0.25]{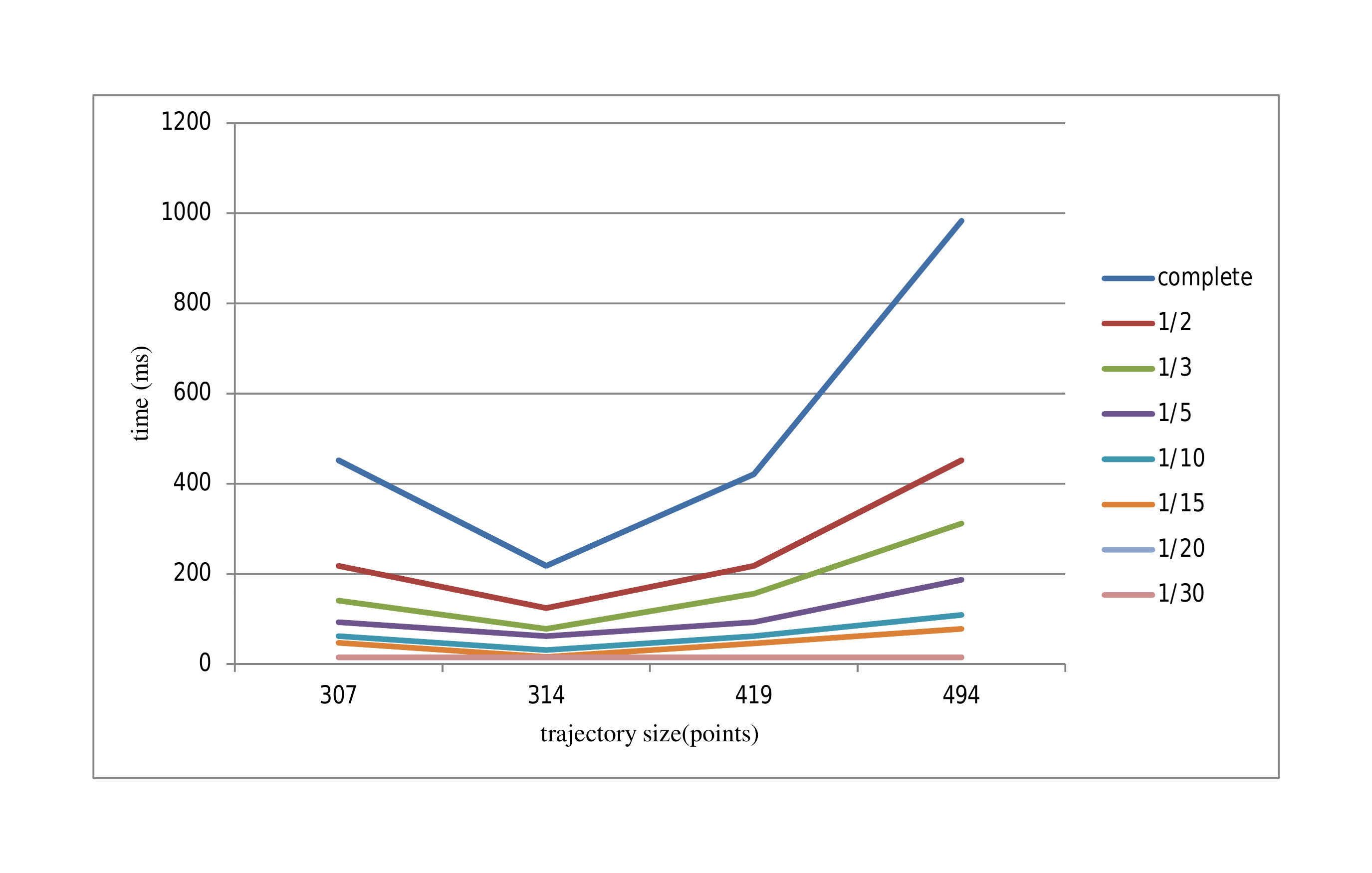}
\caption{Execution time vs. trajectory size}
\label{fig:exetimesizeone}
\end{figure}

\begin{figure}[ht]
\centering
\includegraphics[scale=0.25]{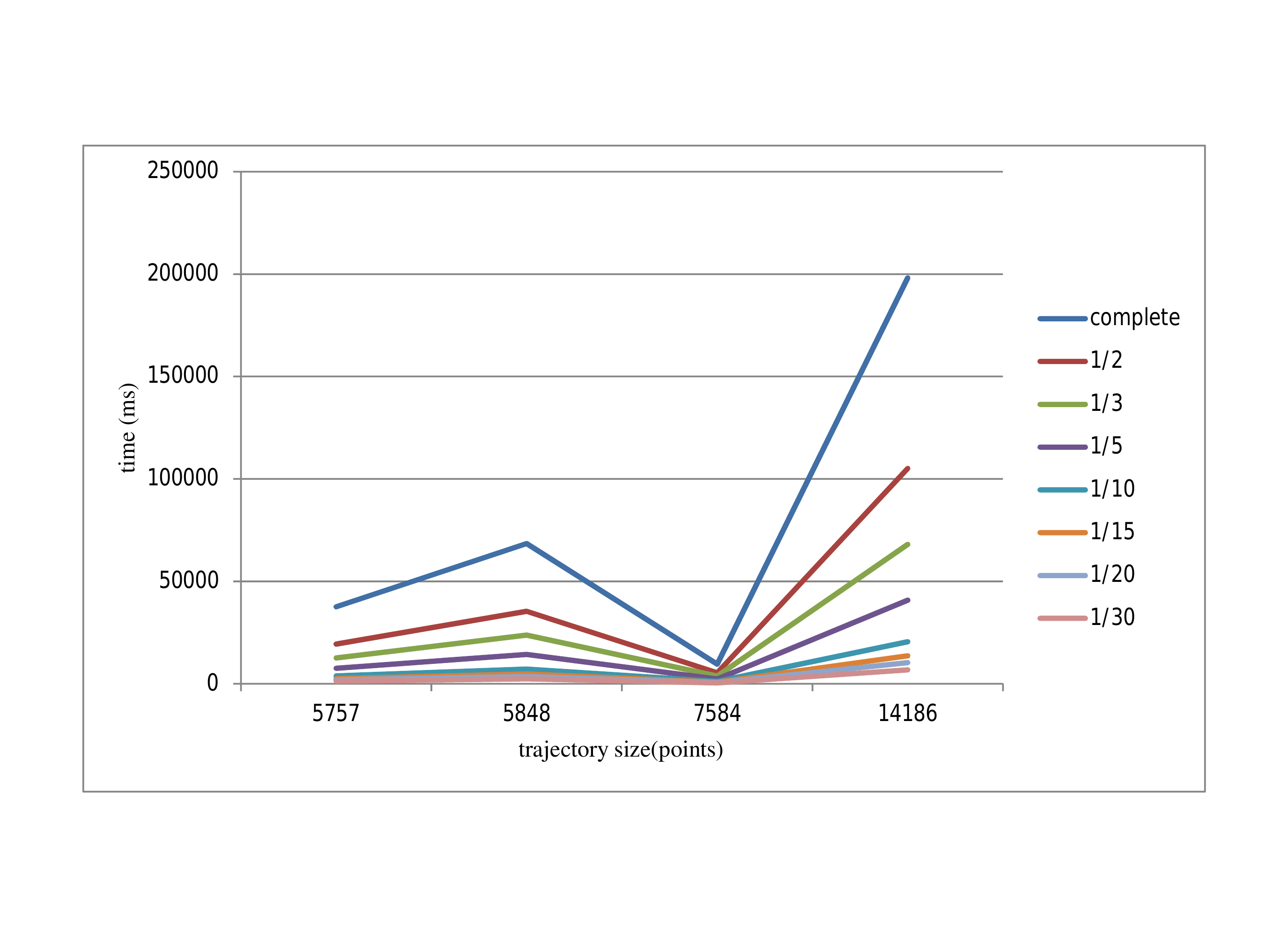}
\caption{Execution time vs. trajectory size}
\label{fig:exetimesizetwo}
\end{figure}

In Figure~\ref{fig:exetimesizeone} and Figure~\ref{fig:exetimesizetwo} one can observe the effect of sample size on execution time for different trajectory sizes. These figures also confirm the aforementioned trade-offs and results.

\section{Open Issues}
\label{sec:openissues}
Although there has been a tremendous amount of work on spatial data quality \cite{xie2008quality,boateng2010assessing} but still many unsolved problems and research possibilities can be found in the context of practical methods for modeling spatial data quality. Moreover, implementing other data quality assessment methods on spatial databases can be an interesting subject for research.

Applying the sample-based method to the spatial data, especially in Iran can be subject of much more research as mentioned in earlier sections.

Formalizing an efficient sampling method, applying sample-based data quality assessment to the spatial data integration of heterogeneous data sources and customizing these methods for spatial data from Iran with multiple aspects and application areas such as natural resources, agriculture and urban design are some of the interesting research areas.

\section{Conclusion}
\label{sec:conclusion}
The importance of spatial data quality is under-estimated especially in Iran and considering the volume of decisions that are made using this type of information, research in this area is essential for credibility of such decisions. Hence we decided to apply one of the most efficient methods for assessing data quality to available spatial data from China and tried to quantify the data quality and to evaluate the accuracy of such data as a base method applicable to other countries. 

Sample-based data quality assessment is an applicable and trustworthy method for estimating quality of spatial data and can be customized and used for deciding whether data available in global or local spatial databases are suitable for using under various circumstances or not. 
Due to the huge volume of spatial data, many spatial operations on them would be very slow and time consuming. In this study we showed that instead of using all of data, we can use samples such that the quality of our operations in many cases is acceptable.

In this approach, samples were collected in a static or periodic manner and because of this unique property, if you are working in a distributed or network environment, the efficiency privilege is considerable and therefore this method is applicable in data quality assessment as well. This will lead to making better decisions in many application domains which convey spatial big data.

\ack
\label{sec X}

\bibliographystyle{unsrtnat}
\bibliography{mainbibio}

\begin{thebibliography}{32}
\providecommand{\natexlab}[1]{#1}
\providecommand{\url}[1]{\texttt{#1}}
\expandafter\ifx\csname urlstyle\endcsname\relax
  \providecommand{\doi}[1]{doi: #1}\else
  \providecommand{\doi}{doi: \begingroup \urlstyle{rm}\Url}\fi

\bibitem[Ghadiri(2006)]{GH}
N.~Ghadiri.
\newblock \emph{Spatial Databases: Overview of Technologies,Techniques and
  Trends}.
\newblock University of Isfahan, October 2006.

\bibitem[Devillers et~al.(2006)Devillers, Jeansoulin,
  et~al.]{devillers2006fundamentals}
Rodolphe Devillers, Robert Jeansoulin, et~al.
\newblock \emph{Fundamentals of spatial data quality}.
\newblock ISTE London, 2006.

\bibitem[Xie et~al.(2008)Xie, Tong, and Jiang]{xie2008quality}
Huan Xie, Xiao-hua Tong, and Zuo-qin Jiang.
\newblock The quality assessment and sampling model for the geological spatial
  data in china.
\newblock \emph{The International Archives of the Photogrammetry, Remote
  Sensing and Spatial Information Sciences, Beijing}, 37:\penalty0 819--824,
  2008.

\bibitem[Goodchild(2007)]{goodchild2007beyond}
M.F. Goodchild.
\newblock Beyond metadata: Towards user-centric description of data quality.
\newblock \emph{ISSDQ}, 2007.

\bibitem[Borek(2011)]{Borek2011Classification}
P.~Borek, A.Woodall.
\newblock A classification of data quality assessment methods.
\newblock \emph{international conference on information quality}, 2011.

\bibitem[Boin and Hunter(2007)]{boin2007communicates}
Anna~T Boin and Gary~J Hunter.
\newblock What communicates quality to the spatial data consumer.
\newblock In \emph{Proceedings of the 2007 International Symposium on Spatial
  Data Quality (ISSDQ 2007), Enschede, The Netherlands}, 2007.

\bibitem[Madnick et~al.(2009)Madnick, Wang, Lee, and Zhu]{madnick2009overview}
Stuart~E Madnick, Richard~Y Wang, Yang~W Lee, and Hongwei Zhu.
\newblock Overview and framework for data and information quality research.
\newblock \emph{Journal of Data and Information Quality (JDIQ)}, 1\penalty0
  (1):\penalty0 2, 2009.

\bibitem[Woodall and Parlikad(2010)]{woodall2010hybrid}
Philip Woodall and Ajith~Kumar Parlikad.
\newblock A hybrid approach to assessing data quality.
\newblock In \emph{Proceedings of the 15th International Conference on
  Information Quality (ICIQ 2010), Little Rock, AR, USA}, 2010.

\bibitem[Peukert et~al.(2011)Peukert, Eberius, and Rahm]{peukert2011amc}
Eric Peukert, Julian Eberius, and Erhard Rahm.
\newblock Amc-a framework for modelling and comparing matching systems as
  matching processes.
\newblock In \emph{Data Engineering (ICDE), 2011 IEEE 27th International
  Conference on}, pages 1304--1307. IEEE, 2011.

\bibitem[Dustdar et~al.(2012)Dustdar, Pichler, Savenkov, and
  Truong]{dustdar2012quality}
Schahram Dustdar, Reinhard Pichler, Vadim Savenkov, and Hong-Linh Truong.
\newblock Quality-aware service-oriented data integration: requirements, state
  of the art and open challenges.
\newblock \emph{ACM SIGMOD Record}, 41\penalty0 (1):\penalty0 11--19, 2012.

\bibitem[Onsrud(2009)]{onsrud2009liability}
Harlan~J Onsrud.
\newblock Liability for spatial data quality.
\newblock 2009.

\bibitem[Talhofer et~al.(2011)Talhofer, Hofmann,
  Ho{\v{s}}kov{\'a}-Mayerov{\'a}, and Kub{\'\i}{\v{c}}ek]{talhofer2011spatial}
V{\'a}clav Talhofer, Alois Hofmann, {\v{S}}{\'a}rka
  Ho{\v{s}}kov{\'a}-Mayerov{\'a}, and Petr Kub{\'\i}{\v{c}}ek.
\newblock Spatial analyses and spatial data quality.
\newblock In \emph{Proceedings of the 14th AGILE International Conference on
  Geographic Information Science-Advancing Geoinformation Science for a
  Changing World, AGILE}, page~8, 2011.

\bibitem[Talhofer et~al.(2009)Talhofer, Hoskova, Kratochvil, and
  Hofmann]{Talhofer2009Geospatial}
V~Talhofer, S~Hoskova, V~Kratochvil, and A~Hofmann.
\newblock Geospatial data quality.
\newblock In \emph{International Conference on Military Technologies
  (ICMT’09)}, pages 570--578, 2009.

\bibitem[Laxmaiah and Govardhan(2013)]{laxmaiah2013conceptual}
M~Laxmaiah and A~Govardhan.
\newblock A conceptual metadata framework for spatial data warehouse.
\newblock \emph{arXiv preprint arXiv:1306.1730}, 2013.

\bibitem[Droj et~al.(2009)Droj, Suba, and Buda]{droj2011}
G~Droj, S~Suba, and A~Buda.
\newblock Modern techniques for evaluation of spatial data quality. revcad.
\newblock \emph{Journal of Geodesy and Cadastre}, \penalty0 (9):\penalty0
  265--272, 2009.

\bibitem[Jakobsson et~al.(2011)Jakobsson, Beare, Marttinen, Onstein, Tsoulos,
  and Williams]{jakobsson2011}
A.~Jakobsson, M.~Beare, J.~Marttinen, E.~Onstein, L.~Tsoulos, and F.~Williams.
\newblock A cohesive approach towards quality assessment of spatial data and
  its automation.
\newblock In \emph{Proceedings of the 25th International Cartographic
  Conference}, pages 33--40, 2011.

\bibitem[M{\'a}rquez et~al.(2012)M{\'a}rquez, Dormann, Sommer, Schmidt,
  Thiombiano, Da, Chatelain, Dressler, and
  Barthlott]{marquez2012methodological}
Jaime R~Garc{\'\i}a M{\'a}rquez, Carsten~F Dormann, Jan~Henning Sommer, Marco
  Schmidt, Adjima Thiombiano, Si{\'e}~Sylvestre Da, Cyrille Chatelain, Stefan
  Dressler, and Wilhelm Barthlott.
\newblock A methodological framework to quantify the spatial quality of
  biological databases.
\newblock \emph{Biodiversity and Ecology}, 4:\penalty0 25--39, 2012.

\bibitem[Devillers et~al.(2007)Devillers, B{\'e}dard, Jeansoulin, and
  Moulin]{devillers2007towards}
Rodolphe Devillers, Yvan B{\'e}dard, Robert Jeansoulin, and Bernard Moulin.
\newblock Towards spatial data quality information analysis tools for experts
  assessing the fitness for use of spatial data.
\newblock \emph{International Journal of Geographical Information Science},
  21\penalty0 (3):\penalty0 261--282, 2007.

\bibitem[Ballou et~al.(2006)Ballou, Chengalur-Smith, and
  Wang]{ballou2006sample}
Donald~P Ballou, InduShobha~N Chengalur-Smith, and Richard~Y Wang.
\newblock Sample-based quality estimation of query results in relational
  database environments.
\newblock \emph{Knowledge and Data Engineering, IEEE Transactions on},
  18\penalty0 (5):\penalty0 639--650, 2006.

\bibitem[Duckham et~al.(2000)Duckham, Drummond, and
  Forrest]{duckham2000spatial}
Matt Duckham, Jane Drummond, and David Forrest.
\newblock Spatial data quality capture through inductive learning.
\newblock \emph{Spatial Cognition and Computation}, 2\penalty0 (4):\penalty0
  261--282, 2000.

\bibitem[Mostafavi et~al.(2004)Mostafavi, Edwards, Jeansoulin,
  et~al.]{mostafavi2004ontology}
Mir-Abolfazl Mostafavi, Geoffrey Edwards, Robert Jeansoulin, et~al.
\newblock An ontology-based method for quality assessment of spatial data
  bases.
\newblock \emph{Third International Symposium on Spatial Data Quality},
  1\penalty0 (28a):\penalty0 49--66, 2004.

\bibitem[Zargar and Devillers(2009)]{zargar2009operation}
Amin Zargar and Rodolphe Devillers.
\newblock An operation-based communication of spatial data quality.
\newblock \emph{Advanced Geographic Information Systems Web Services
  International Conference}, pages 140--145, 2009.

\bibitem[Vizhi and Bhuvaneswari(2012)]{vizhi2012data}
J~Vizhi and T~Bhuvaneswari.
\newblock Data quality measurement on categorical data using genetic algorithm.
\newblock \emph{arXiv preprint arXiv:1202.3215}, 2012.

\bibitem[Henriksson(2007)]{Henriksson2007Ontology}
T.~Henriksson, R.M.Kauppinen.
\newblock An ontology-driven approach for spatial data quality evaluation.
\newblock \emph{The International Archives of the Photogrammetry, Remote
  Sensing and Spatial Information Sciences}, \penalty0 (34), 2007.

\bibitem[Hunter(1999)]{hunter1999new}
Gary~J Hunter.
\newblock New tools for handling spatial data quality: moving from academic
  concepts to practical reality.
\newblock \emph{URISA Journal}, 11\penalty0 (2):\penalty0 25--34, 1999.

\bibitem[Chen et~al.(2009)Chen, Jiang, Zheng, Li, and
  Yu]{DBLP:conf/gis/ChenJZLY09}
Yukun Chen, Kai Jiang, Yu~Zheng, Chunping Li, and Nenghai Yu.
\newblock Trajectory simplification method for location-based social networking
  services.
\newblock In Xiaofang Zhou and Xing Xie, editors, \emph{Proceedings of the 2009
  International Workshop on Location Based Social Networks, {LBSN} 2009,
  November 3, 2009, Seattle, Washington, USA, Proceedings}, pages 33--40.
  {ACM}, 2009.
\newblock \doi{10.1145/1629890.1629898}.
\newblock URL \url{http://doi.acm.org/10.1145/1629890.1629898}.

\bibitem[Levy and Lemeshow(2013)]{levy2013sampling}
Paul~S Levy and Stanley Lemeshow.
\newblock \emph{Sampling of populations: methods and applications}.
\newblock John Wiley and Sons, 2013.

\bibitem[Hunter et~al.(2005)Hunter, Hope, Sadiq, Boin, Marinelli, Kealy,
  Duckham, and Corner]{hunter2005next}
GJ~Hunter, S~Hope, Z~Sadiq, A~Boin, M~Marinelli, AN~Kealy, M~Duckham, and
  RJ~Corner.
\newblock Next-generation research issues in spatial data quality.
\newblock 2005.

\bibitem[Devillers et~al.(2010)Devillers, Stein, B{\'e}dard, Chrisman, Fisher,
  and Shi]{devillers2010thirty}
Rodolphe Devillers, Alfred Stein, Yvan B{\'e}dard, Nicholas Chrisman, Peter
  Fisher, and Wenzhong Shi.
\newblock Thirty years of research on spatial data quality: achievements,
  failures, and opportunities.
\newblock \emph{Transactions in GIS}, 14\penalty0 (4):\penalty0 387--400, 2010.

\bibitem[Guptill(2008)]{DBLP:journals/tgis/Guptill08}
Stephen~C. Guptill.
\newblock \emph{Fundamentals of Spatial Data Quality} - edited by {Rodolphe}
  {Devillers} {And} {Robert} {Jeansoulin}.
\newblock \emph{T. {GIS}}, 12\penalty0 (1):\penalty0 161--162, 2008.
\newblock \doi{10.1111/j.1467-9671.2008.01091.x}.
\newblock URL \url{http://dx.doi.org/10.1111/j.1467-9671.2008.01091.x}.

\bibitem[Agumya and Hunter(1999)]{hunter1999risk}
A.~Agumya and G.J. Hunter.
\newblock A risk-based approach to assessing the fitness for use of spatial
  data.
\newblock \emph{URISA}, 1\penalty0 (11), 1999.

\bibitem[Boateng(2010)]{boateng2010assessing}
I.~Boateng, B.~Yakubu.
\newblock Assessing the quality of spatial data.
\newblock \emph{European Journal of Scientific Research}, 43\penalty0
  (4):\penalty0 507--515, 2010.

\end{thebibliography}

\vfill\eject

\footnotesize

\noindent\begin{minipage}{\columnwidth}
\begin{wrapfigure}{l}{1.2cm}
  \begin{center}
    \includegraphics[width=2cm]{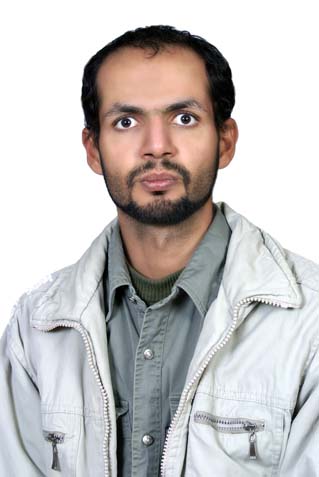}
  \end{center}
\end{wrapfigure}
\noindent
\newline
\textbf{Bagher Saberi} recieved a Bachelor of Science degree in hardware engineering from Ferdowsi University of Mashhad, Mashhad, Iran, 2010.He received a Master of Science degree in Computer Engineering from Isfahan University of Technology, Isfahan, Iran, 2014. He is Currently working as Network Administrator in National Iranian Copper  industries Company.
\end{minipage}


\end{document}